\documentstyle[prd,aps,multicol,epsfig,eqsecnum,amssymb]{revtex}
\def\bbbone{{\mathchoice {\rm 1\mskip-4mu l} {\rm 1\mskip-4mu l}
{\rm 1\mskip-4.5mu l} {\rm 1\mskip-5mu l}}}

\begin{document}

\title{Local D=4 Field Theory on $\kappa$--Deformed Minkowski
Space\footnote{Supported by KBN grant 2P03B13012}}

\author{P. Kosi\'{n}ski} 
 
\address{Institute of Physics, 
University of L\'{o}d\'{z}, \\
ul. Pomorska 149/53 90--236 L\'{o}d\'{z}, Poland}

\author{J. Lukierski}
\address{Institute for Theoretical Physics,\\
 University of Wroc{\l}aw,
pl. Maxa Borna 9,\\
 50--205 Wroc{\l}aw, Poland
\\
and
\\
Departamento de Fisica Teorica and IFIC (CSIC) \\
Facultad de Fisica, Universidad de Valencia\\
46100 Burjasot (Valencia), Spain}

\author{P. Ma\'{s}lanka}
\address{Institute of Physics, 
University of L\'{o}d\'{z}, \\
ul. Pomorska 149/53 90--236 L\'{o}d\'{z}, Poland}

\date{\today}
\maketitle

\begin{abstract}
We describe the local D=4 field theory on $\kappa$--deformed Minkowski
space as nonlocal relativistic field theory on standard Minkowski
space--time. For simplicity the case of $\kappa$-deformed scalar
field $\phi$ with the interaction 
$\lambda \phi^{4}$ is considered, and the $\kappa$--deformed
interaction vertex is described. It appears that fundamental
mass parameter $\kappa$ plays a role of regularizing imaginary 
Pauli--Villars mass in $\kappa$--deformed  propagator. 

\end{abstract}

\begin{multicols}{2}

\section*{Introduction}
The conventional field--theoretic models in elementary particle
physics (e.g. standard model -- Yang-Mills theory with spinorial
and Higgs sector) are described by quantum field theory on standard
D=4 Minkowski space. The addition of gravity effects can be
realized in the following two steps:

a) We consider field--theoretic models in curved background of
classical gravitational field, describing curved space--time geometry.

b) Further we quantize curved space--time geometry by considering
quantized gravitational field.

At present it is not known how to handle with reasonable
accuracy the quantum gravity modification of geometry, e.g. describe field
theories in the background of quantized geometries.
It is known moreover, that due to divergent quantum effects in
Einstein gravity the notion of classical space--time can not be
used at  distances comparable and smaller than the Planck
length $l_{p}$ ($l_{p}\simeq 10^{-33}$cm) (see e.g. [1-3]). 
In front of this
difficulty one way of generalizing the standard framework of local
relativistic fields is to replace classical space--time points
by some primary extended objects (e.g. fundamental strings,
p-branes etc.) providing new field theory (e.g. string field
theory) which leads to 
 finite quantum corrections. Another way, closer to 
  the idea of quantized
geometry, is to replace commuting space--time coordinates by
noncommuting generators of a quantized Minkowski space. There is a hope
that such a quantized space--time geometry will provide additional
convergence factors or even finite quantum field theory.
Indeed, as we shall show below, if we introduce mass--like
deformation parameter $\kappa$, it occurs also as a regularizing
 imaginary 
Pauli--Villars large mass parameter, describing tachyonic pole
in the $\kappa$--deformed propagator.

\section{$\kappa$--deformed $D=4$ relativistic symmetries }
The standard space--time coordinates $x_{\mu}=(\vec{x},x_{0})$
can be described as translations which form the Abelian subgroup
of $D=4$ Poincar\'{e} group ${\cal{P}}_{4}$ (formally one can
identify the Minkowski space with the coset
${\cal{P}}_{4}/O(3,1)$, where $O(3,1)$ describes the Lorentz
subgroup). Similarly the properties of four relativistic momenta
can be described by the four translation generators of $D=4$
Poincar\'{e} algebra. In such a scheme one considers the
relativistic Poincar\'{e} symmetries, described by the dual 
pair - $D=4$ Poincar\'{e} group and $D=4$ Poincar\'{e} algebra - as
primary geometric notions
from which the properties of Minkowski
space, fourmomenta as well as relativistic phase space are derived. This
primary status of symmetries we shall keep also in the case when
the classical relativistic framework is modified by the
procedure of quantum deformations.
In analogy with classical case, the quantum Minkowski space and quantum
fourmomentum space can be obtained, respectively,
from the  translation sector of D=4 quantum--deformed
Poincar\'{e} group and fourmomentum generators belonging to quantum
Poincar\'{e} algebra.

In the last 
 years  there were proposed different quantum deformations of
D=4 Poincar\'{e} symmetries (see e.g.  [4-7])
 in the form of real noncommutative and noncocommutative 
Hopf algebras.
In particular, as a first such deformation there was proposed
 the so--called $\kappa$--deformation  
 [4,5]
introducing fundamental mass (or length) parameter $\kappa$. 
The $\kappa$--deformed Poincar\'{e}
 algebra 
  [4,8] is the associative and coassociative Hopf algebra with generators
$M_{\mu\nu} = (M_{i}={1\over 2} \epsilon_{ijk}M_{jk},
N_{i}=M_{i0}$
$P_{k}=(P_{i},P_{0})$ 
satisfying the relations\footnote{The $\kappa$--deformed algebra
can be written with the use of different basic generators. The
one presented below uses so--called bicrossproduct  basis [8]
with the  quantum deformation in the algebraic sector
 occurring entirely in 
the ``cross" commutation relations of the Abelian fourmomentum
generators with  the classical Lorentz generators.}

\begin{mathletters}
\label{1a}
\begin{eqnarray} 
 [M_{\mu\nu},M_{\rho\tau}] 
=&&
\displaystyle
i\left(\eta_{\mu\tau}M_{\nu\rho} -
\eta_{\nu\rho}M_{\nu\tau} +\eta_{\nu\rho}M_{\mu\tau}
\right.
\cr\cr
&&
\left.
  -\  \eta_{\nu\tau}M_{\mu\rho} \right),
\cr \cr
 [M_{i},P_{j}] =&&  i\epsilon_{ijk}P_{k}\, , 
\quad [M_{i},P_{0}]= iP_{i}\, ,
\cr \cr
\displaystyle
  [N_{i}, P_{j}] = && i\delta_{ij}
\left[ \left( {\kappa\over 2} \left(
 \bbbone -e^{-2{P_{0}/ \kappa}}
\right) + {1\over 2\kappa} \vec{P}\, ^{ 2}\, \right)
\right.
\cr\cr
&&
\left.
- \ {i\over \kappa} P_{i}P_{j}\right] ,
\cr \cr
 \left[N_{i},P_{0}\right] = && iP_{i}\, ,
\cr \cr
\displaystyle
  [P_{\mu},P_{\nu}] = && 0\, , 
\label{1.1a}
\end{eqnarray} 

with the coproducts, antipodes and counits defined by

\begin{eqnarray} 
\displaystyle
 && \Delta(M_{i}) = M_{i}\otimes \bbbone + \bbbone
\otimes M_{i}\, ,
\cr\cr
\displaystyle
&& \Delta(N_{i}) = N_{i}\otimes e^{-{P_{0}/ \kappa}} +
\bbbone \otimes N_{i} - {1\over \kappa}
\epsilon_{ijk}M_{j}\otimes P_{k}\, ,
\cr\cr
\displaystyle
&& \Delta(P_{i}) = P_{i}\otimes e^{-{P_{0}/ \kappa}} +
\bbbone \otimes P_{i}\, ,
\cr\cr
\displaystyle
&& \Delta(P_{0}) = P_{0}\otimes \bbbone +  \bbbone \otimes P_{0}\, ,
\label{1.1b}
\end{eqnarray}

 \begin{eqnarray} 
&&  S(M_{i}) = - M_{i}\, , 
\cr\cr
&& S(N_{i})= - \left( N_{i} + {1\over \kappa} \epsilon_{ijk}
 M_{j}P_{k}\right) e^{P_{0}/ \kappa}\, ,
\cr\cr
&&  S(P_{i}) = -P_{i}e^{P_{0}/ \kappa}  
\cr\cr
&& 
S(P_{0}) = -
P_{0}\, ,
\label{1.1c}
\end{eqnarray}

 \begin{eqnarray}
&& \epsilon(P_{\mu},M_{i},N_{i}) = 0\, . 
 \end{eqnarray}
\label{1.1d}
\end{mathletters}
The classical Poincar\'{e} algebra is obtained in the 
 limit $\kappa  \to \infty$. 

The Hopf algebra  (1.1a--d) has  its dual form, 
 generated by
the following duality relations
\begin{mathletters}
\label{1.2a}
\begin{eqnarray}
&&\langle P^{\mu}, \hat{x}_{\rho}\rangle = \delta ^{\mu}_{\ \rho}
\\ \cr
&&\langle M^{\mu\nu}, \Lambda_{\rho \tau}\rangle =
\delta^{\mu}_{\ \rho} \delta^{\nu}_{\ \tau} -
\delta^{\mu }_{\ \tau}\delta^{\nu}_{\ \rho}
\end{eqnarray}
\end{mathletters}
In accordance with general scheme the dual generators
$(\hat{x}_{\rho}, \Lambda_{\rho \tau})$ describe
$\kappa$--deformed Poincar\'{e} group. Using duality relations
between arbitrary powers of the generators
$(P^{\mu},M^{\mu\nu})$ and $(x_{\rho},\Lambda_{\rho\tau})$ 
 as well as
the property of inner product
\begin{eqnarray}
&\langle A \otimes B, \Delta(a)\rangle &= \langle A \cdot B, a\rangle
\cr\cr
& \langle \Delta(A), a\otimes b\rangle  & =
\langle A, a \cdot b \rangle
\end{eqnarray} 
one can derive the following set of relations, defining
$\kappa$--deformed Poincar\'{e} group [5,8]:

 a) algebraic relations 
\begin{mathletters}
\begin{eqnarray} 
 \left[ \hat{x}^{\mu}, \hat{x}^{\nu}\right]
&=& {i \over \kappa} \left( \delta_{0}^{\mu} \hat{x}^{\nu}
- \delta_{0}^{\ \nu} \hat{x}^{\mu}\right)\, ,
\cr\cr
\left[ {\Lambda}_{\ \nu}^{ \mu}, \Lambda^{\rho}_{\ \tau}\right]
&=& 0\, ,
\cr\cr
\left[ {\Lambda}_{\ \nu}^{ \mu}, \hat{x}^{\rho}\right]
&=& - 
{1\over \kappa} 
\left[ \left( \Lambda^{\mu}_{\ 0} - \delta_{0}^{\ \mu}\right)
\Lambda^{\rho}_{\ \nu}
\right.
\cr\cr
&&
\left.
 +
\left( \Lambda^{0}_{\nu} - \delta^{\ 0}_{\nu}\right)
\eta^{\mu\rho} \right]\, ,
\label{1.2b}
\end{eqnarray}

b) coproducts

\begin{eqnarray} 
 \Delta(\hat{x}^{\mu}) &=&
\Lambda^{\mu}_{\ \rho} \otimes \hat{x}^{\rho} + \hat{x}^{\mu} \otimes
\bbbone\, ,
\cr\cr
 \Delta(\Lambda^{\mu}_{\  \nu}) &=& \Lambda_{\ \rho}^{\mu}
\otimes \Lambda^{\rho}_{\ \nu } \, ,
\end{eqnarray}

c) antipodes and counits

\begin{eqnarray} 
& S(\Lambda^{\mu}_{\ \nu} ) =- \Lambda_{\ \nu}^{ \mu}\, ,
\quad &
S(x^{\mu})= - x^{\mu}\, ,
\cr\cr
& \epsilon(\Lambda^{\mu}_{\ \nu}) = \delta^{\mu}_{\ \nu}\, ,
\quad &
\epsilon(x^{\mu})= 0\, .
\end{eqnarray}
\end{mathletters}
It should be observed that the relations (1.2b) and (1.2c)
remains undeformed, as for classical Poincar\'{e} group.

Our aim in this paper is to consider the field theory on
$\kappa$--deformed Minkowski space, described by the
translation generators $\hat{x}_{\mu}$ of the $\kappa$--deformed
Poincar\'{e} group (see (1.4a--c)) after the contraction mapping
$\Lambda^{\mu}_{\ \alpha} \to \delta^{\mu}_{\ \alpha}$.

 The $\kappa$--deformed Hopf algebra $H_{x}$  describing the
  quantum
 Minkowski space--time is generated by the coordinates  $\hat{x}_{\mu}$ 
 and determined by the following basic relations:

\begin{mathletters}
\begin{eqnarray}
&& 
 [\hat{x}_{0},\hat{x}_{i}] = {i\over \kappa} \, \hat{x}_{i}\, ,
\qquad
[\hat{x}_{i},\hat{x}_{j}] = 0\, ,
\label{1.5a}
\\  \cr
&& 
\Delta(\hat{x}_{\mu}) = 
\hat{x}_{\mu}\otimes \bbbone + \bbbone \otimes \hat{x}_{\mu}\, ,
\end{eqnarray}
\end{mathletters}
The dual Hopf algebra $H_{p}$ of functions on $\kappa$--deformed
fourmomenta is described by the Hopf subalgebra of the
$\kappa$--deformed Poincar\'{e} algebra  (1.1a--c) 
 as follows
\begin{mathletters}
\begin{equation}
 [p_{\mu}, p_{\nu} ] = 0\, ,
\label{1.6a}
\end{equation}
\begin{eqnarray}
 \Delta(p_{i}) &=& p_{i} \otimes e^{-{P_{0}/ \kappa}}
+ \bbbone \otimes p_{i}\, ,
\cr\cr
 \Delta(p_{0}) &= & p_{0} \otimes \bbbone 
+ \bbbone \otimes p_{0}\, .
\end{eqnarray}
\end{mathletters}
The antipodes and counits in $ H_{x}$ and $H_{p}$ remain   classical.

An important tool will be the use of $\kappa$--deformed Fourier
transform, describing the fields on non--commutative Minkowski
space with generators $\hat{x}=(\hat{x}_{i},\hat{x}_{0})$
in the following way $(p\hat{x}\equiv p_{i}\hat{x}_{i} -
p_{0}\hat{x}_{0})$\footnote{One can also use in (1.7) the
measure $\Omega^{4} =d^{4}p \, e^{- {3p_{0}\over \kappa}}$ which
is invariant under the shift in fourmomentum space described by
the coproduct (1.6b).}

\begin{equation}
\Phi(\hat{x}) = {1\over (2\pi)^{4}}
\int d^{4}p\  \widetilde{\Phi} (p):\, e^{ip\hat{x}}\, ;
\end{equation}
 $\widetilde{\phi}(p)$ is a classical function on commuting
fourmomentum space $p=(p_{i}, p_{0})$ and
 $(\vec{p}\,  \vec{x}\,  \equiv p_{i}\hat{x}_{i})$

\begin{equation}
 :e^{ip\hat{x}}:\,  
 \doteq
 e^{ip_{0}\hat{x}_{0}}
 e^{i\vec{p}\, \vec{x}}\, .
 \end{equation}
 
 The $\kappa$--deformed exponential (1.4b) describes the canonical element
  [9,10] or $T$--matrix [11,12]
  for the pair of dual Hopf algebras $H_{x}$ and $H_{p}$.

From (1.6b) and (1.8) follows that
\begin{equation}
:\, e^{ip\hat{x}} \, :\, : \, e^{iq\hat{x}} \, : \,
= \, :\, 
e^{i\Delta^{(2)} (p,q)\hat{x}}\, : 
\label{1.9}
\end{equation}
where $\Delta^{(2)}(p,q) = (\vec{p}\, e^{-{q_{0}/ \kappa}}
+ \vec{q}, p_{0} + q_{0} )$\ is the fourmomentum addition rule
described by the coproduct~(1.6b).

From the Fourier transform $\widetilde{\Phi}(p)$ one can obtain also
a standard relativistic field $\phi(x)$ on classical Minkowski
space $x\equiv (x_{i},x_{0})$ by performing  the classical
Fourier transform
\begin{equation}
{\phi}(x) = {1\over (2\pi)^{4}}
\int d^{4}p \ {\widetilde{\Phi}}(p) \, e^{ipx}\, .
\label{1.10}
\end{equation}

It is easy to see that in the limit $\kappa \to \infty$ we get
$\hat{x}_{\mu} \to x_{\mu}$ and the formula (1.7) describing
``quantum" Fourier transform  can be
identified with the classical
 one, given by (1.10).

We shall describe the $\kappa$--deformation of relativistic
local field theory in the following three steps:

1) We replace in conventional local relativistic field theory
the classical Minkowski coordinates $x_{\mu}$ by quantum
Minkowski coordinates $\hat{x}_{\mu}$, and
relativistic--covariant differential operators defining free
fields by corresponding $\kappa$--covariant differential
operators on $\kappa$--deformed Minkowski space. 
The  $\kappa$--deformed Lagrangean is the product of fields
(1.7) and its derivatives and becomes 
  an element of noncommutative Hopf algebra defined by (1.5).
The new action
is obtained by integration of local products of
$\kappa$--deformed fields and their derivatives over
$\kappa$--deformed Minkowski space.

2) In order to perform the integration we
  substitute in the $\kappa$--deformed Lagrangean the Fourier
transforms (1.7), apply the formula (1.9), and replace the
integration over $\kappa$--deformed Minkowski space by the
$\kappa$--deformed convolution integrals in fourmomentum space. 
 We can calculate all
occurring $\kappa$--deformed integrals $\int\!\!\!\int d^{4}\hat{x}$ by
using the formula

\begin{equation}
{1 \over (2\pi)^{4}}
\int\!\!\!\!\!\int d^{4} \hat{x}\, : e^{ip\hat{x}} \, : \, =
\delta^{4}(p)\, .
\end{equation}
 which implies for example that
\begin{mathletters}
\begin{eqnarray}
&&\int\!\!\!\!\!\int \phi(\hat{x})d^{4}x = \int d ^{4}p \delta^{4}(p)
\widetilde{\Phi}(p) = \widetilde{\Phi}(0)
\\ \cr
&&\int\!\!\!\!\!\int \phi^{2}(\hat{x})d^{4}x 
\cr\cr
&&= 
\int d ^{4}p_{1} \int d^{4}p_{2}
\widetilde{\Phi}(p_{1}) 
 \widetilde{\Phi}(p_{2})
 \delta(\Delta^{(2)} (p_{1},p_{2})) 
 \cr\cr
 &&=
 \int d^{4}p\widetilde{\Phi} \left(\vec{p},p_{0}\right)
 \left(-\vec{p}\, e^{P_{0}/\kappa}, -p_{0}\right)
 \end{eqnarray}
 and in general case
 \begin{eqnarray}
 && \int\!\!\!\!\!\int d^{4}\hat{x}
  \Phi^{n}(\hat{x})
   = \int d ^{4}p^{(1)} \int d^{4}p^{(n)}
  \Phi(p^{(1)} )
  \cr\cr 
  && \quad
  \ldots \Phi(p^{(n)})
  \delta^{(4)}\left(\Delta^{(n)}_{\mu} 
  \left( p^{(1)}, \ldots p^{(n)}\right)\right)
  \end{eqnarray}
  \end{mathletters}
  where $ \Delta^{(n)}$ is the iterated coproduct (1.4b)
  \begin{eqnarray}
&&  \Delta_{0}^{(n)}\left( p^{(1)}, \ldots p^{(n)} \right)
  =  p^{(1)}_{0} + \ldots +  p^{(n)}_{0}
  \cr\cr
  &&  \vec{\Delta}\, ^{ (n)}\left( \vec{p}\, ^{ (1)},
   \ldots \vec{p}\, ^{ (n)} \right)
= \sum\limits^{n}_{k=1} \vec{p}\, ^{ (k)} \, e^{-{ 1\over \kappa}
\sum\limits^{n}_{l=k+1} p^{(l)}_{0}} 
\end{eqnarray}
The integral (1.12a) in the two-dimensional case in different context
has been used by Majid [13,14].

3) In order to interprete the $\kappa$--deformation in standard
Minkowski space we use the formula (1.10). The $\kappa$--deformed
convolution integrals of Fourier transforms $\widetilde{\phi}(p)$
become the $\kappa$--dependent nonlocal vertices of fields
$\phi(x)$ in standard Minkowski space.

\section{$\kappa$--deformed Minkowski space: differential
bicovariant calculus, vector fields and integration}

In order to describe the field equations on $\kappa$--deformed
Minkowski space one should consider corresponding differential
calculus and its covariance properties  under the action of
$\kappa$--deformed Poincar\'{e} group.

{\bf a) Differential bicovariant calculus.}

On $\kappa$--deformed $H_{\kappa}$ with four selfadjoint
generators $\hat{x}_{\mu}$ one can construct a five--dimensional
bicovariant differential calculus with the basis [15,16,17]
\begin{equation}
\tau^{\mu} = d\hat{x}^{\mu}\, , \qquad
\tau^{5}= [d\hat{x}^{\mu},\hat{x}_{\mu}]
+ {3i\over \kappa} d\hat{x}^{0}\, ,
\end{equation}
satisfying the relations

\begin{eqnarray}
\displaystyle
 [\tau^{\mu},\hat{x}^{\nu}] &=&
{i\over \kappa} \eta^{0 \mu} \tau^{\nu}
- {i\over \kappa} \eta^{\mu\nu} \tau^{0} +
{1\over 4} \eta^{\mu\nu} \tau\, ,
\cr\cr
\displaystyle
 [\tau^{5},\hat{x}^{\nu}] & = &  - {4\over \kappa^{2}}
\tau^{\nu}\, ,
\end{eqnarray}

and $(A=0,1,2,3,4,5)$

\begin{eqnarray}
&&
\tau^{A}\wedge \tau^{B} = -\tau^{A} \wedge \tau^{B}\, ,
\cr\cr
&&
d\tau^{\mu} = 0 \qquad d\tau^{5} = - 2\tau^{\mu}
\wedge \tau_{\mu}\, .
\end{eqnarray}

The $\kappa$--Minkowski space (1.5a-b) carries the left
covariant action of $\kappa$--Poincar\'{e} group
(1.4a-c)

\begin{equation}
\rho_{L}(\hat{x}^{\mu}) = \Lambda^{\mu} _{\ \nu}
\otimes \hat{x}^{\nu} + a^{\mu}\otimes \bbbone\, ,
\end{equation}
and the relations (2.2-3) are covariant under the following
transformations of differentials
\begin{equation}
\widetilde{\rho}_{L}(\tau^{\mu}) =\Lambda^{\mu}_{\ \nu}
\otimes \tau^{\nu}\, ,
\qquad
\widetilde{\rho}_{L}(\tau^{5}) = \bbbone \otimes \tau^{5}\, .
\end{equation}
One can also define the covariant right action of another Poincar\'{e}
group
\begin{equation}
\rho_{R}(\hat{x}^{\mu}) = \hat{x}^{\nu} \otimes
\Lambda_{\nu}^{\ \mu} - \bbbone \otimes a^{\nu}\Lambda^{\
\mu}_{\nu}\, ,
\end{equation}
obtained by the change $\kappa \to -\kappa$ in the formulae
(1.2). Such right covariance quantum group provides the
following covariance of the relations (2.2-3)
\begin{equation}
\widetilde{\rho}_{L} (\tau^{\mu}) = \tau^{\nu}\otimes
\Lambda_{\nu}^{\ \mu}\, ,
\qquad \widetilde{\rho}_{L}(\tau^{5})=\tau^{5}\otimes \bbbone\, ,
\end{equation}

{\bf b) $\kappa$--deformed vector fields.}

In order to define the vector fields describing left or right
partial derivatives $\hat{\partial}_{A}$ acting on functions on
$\kappa$--deformed Minkowski space we write:
\begin{equation}
df = \hat{\partial}_{A}^{L} \, f\cdot \tau^{A} =
\tau^{A} \, \hat{\partial}_{A}^{R}\, f\, .
\end{equation}
In particular, using Leibnitz rule $d(fg)=dfg+fdg$ we obtain
that\footnote{Further we shall use the left partial derivatives
and put $\hat{\partial}_{A}\equiv \hat{\partial}_{A}^{L}$.}
\begin{equation}
d\, : \, e^{ip\hat{x}} \,: \, = \, : \,
\chi_{A}(p_{\mu})e^{ip\hat{x}} \, : \, \tau^{A}\, ,
\end{equation}
where

\begin{eqnarray}
\chi_{i} &= & e^{P_{0}/ \kappa} \, p_{i}\, ,
\cr\cr
\chi_{0} & =&  \kappa\left( e^{P_{0}/ \kappa} - 1\right)
+ {1\over 2\kappa} \, M_{\kappa}^{2} (p)\, ,
\cr\cr
\chi_{5} & =  & -{1\over 8} M_{\kappa}^{2}(p) 
\end{eqnarray}
and
\begin{eqnarray}
M^{2}_{\kappa}(p) &=&  e^{P_{0}/ \kappa} \, \vec{p}\, ^{2} -
\left( 2\kappa \ \sinh {P_{0}\over 2\kappa} \right)^{2}
\cr\cr
& = &  e^{P_{0}/ \kappa} \,  \vec{p}\, ^{2} + 2\kappa^{2}
\left( 1 - \cosh  {P_{0}\over \kappa} \right)
\, ,
\end{eqnarray}
where in the limit $\kappa  \to \infty$ we get 
 $M^{2}_{\kappa}(p) \to  \vec{p}\, ^{2} - p^{2}_{0} = -m^{2}$. 
 Using  Fourier transform
(1.3) and the relation 
$p_{\mu}:e^{ip\hat{x}}:=:{1\over
i}\partial_{\mu}e^{ip\hat{x}}:$ 
 one obtains
\begin{equation}
\partial_{A}\Phi(\hat{x}) = \, :\,
\chi_{A}\left({1\over i}{\partial\over \partial x^{\mu}}\right)
\Phi(\hat{x})\, :
\end{equation}
Using duality relations one can  relate the derivatives on $
\kappa$--deformed Minkowski space with the fourmomentum
generators (1.4a--b).
In bicrossproduct basis one obtains [8]
\begin{equation}
\langle P_{\mu}f(P), \, : \,
\phi(\hat{x})\, :\rangle =
\langle f(P), \, : \,
 i{\partial \over \partial x^{\mu}}
 \Phi(\hat{x})\rangle\, ,
 \end{equation}
 where $P_{\mu} \in H_{p}$, or equivalently
 \begin{equation}
 P_{\mu}\, : \, \phi(\hat{x}) \, : \, =
 \, : \, {1\over i} {\partial\over \partial x^{\mu}}
 \phi(\hat{x})\, :
 \end{equation}
We see that one can express the vector fields in terms of the fourmomentum
generators $P_{\mu}$ of $\kappa$--deformed Poincar\'{e} algebra (1.1):
\begin{equation}
\partial_{A}\Phi(\hat{x}) =
\chi_{A}(P_{\mu}): \phi(\hat{x})\, :
\end{equation}
where $\chi_{A}$ are given by relations (2.10).

{\bf c) $\kappa$--invariant integration over $\kappa$--deformed Minkowski
space.}

The relation (1.12a) and the definition (1.8) lead
to the equality\footnote{The idea that integration over
noncommutative space of suitably ordered function is equal to
the classical integral first appeared in [13,17].}
\begin{equation}
{1\over (2\pi)^{4}}\int\!\!\!\!\!\int d^{4}\hat{x}\ 
\Phi (\hat{x}) =
\widetilde{\Phi}(0) =
{1\over (2\pi)^{4}} \int d^{4} x \ \phi(x)\, .
\end{equation}

In order to show the $\kappa$--Poincar\'{e} invariance of (2.16)
we should prove that\footnote{We shall use the right action of
the $\kappa$--deformed Poincar\'{e} symmetry.}
\begin{equation}
\int\!\!\!\!\!\int d^{4}\hat{x}
\Phi\left( \hat{x}^{\mu} \otimes 
\Lambda^{\ \nu}_{\mu} -
I \otimes a^{\mu}\Lambda^{\ \nu}_{\mu}\right)
=\widetilde{\Phi}(0) \otimes \bbbone\, .
\end{equation}
The relation (2.17) is shown in the Appendix.

It is easy to see from (1.12b--d) that the relation (2.16) can
not be generalized to the powers of functions; in particular
\begin{equation}
{1\over (\pi)^{4}} 
\int\!\!\!\int 
d^{4}\widehat{x} \phi^{n}(\widehat{x}) \neq
{1\over (\pi)^{4}} \int d^  {4}x \phi^{n}(x) \quad n\geq 2 
\end{equation}

{\bf d) Hermitean conjugation, adjoint derivatives
$\hat{\partial}^{+}_{A}$. }

If we denote
\begin{equation}
\Phi^{+}({\hat{x}}) = 
{1\over(2\pi)^{4}} \int d^{4}p \, 
{\widetilde{\Phi}}^{+}({p}): e^{ip\hat{x}}:
\end{equation}
one easily calculates that
\begin{equation}
{\widetilde{\Phi}}^{+}(\vec{p},p_{0})
=
e^{- {3p_{0}/ \kappa}}
{\widetilde{\Phi}}
\left( 
- e^{p_{0}/ \kappa} \vec{p}, -p_{0} 
\right)\, .
\end{equation}
The Fourier transform (2.19) substituted in (1.10) provide the
following notion of $\kappa$--deformed adjoint operation in
standard Minkowski space:
\begin{equation}
\phi^{+} (x)
= e^{- {i\partial_{t}/ \kappa}
(3+\vec{x}\, \vec{\nabla}\, ) }
\phi^{\star}(x)
= \hat{R}_{\kappa}
\phi^{\star}
\left(
\vec{x}, x_{0} 
- {3i\over 2\kappa} \right)\, ,
\end{equation}
where $R_{\kappa}= \exp{1\over \kappa}\partial_{0}{\cal D}$
and ${\cal D} = -i\left( 
\vec{x}\, \vec{\nabla}\, +{3\over 2}
\right)$. 
The formula (2.19) also defines the notion of adjoint
derivatives, satisfying the relation corresponding to the
integration by parts:
\begin{equation}
 \int\!\!\!\!\!\int d^{4}\hat{x}
\Phi_{1}(\hat{x}) 
\left(
\hat{\partial}_{A} 
\Phi_{2}(\hat{x})
\right)
 =
\int\!\!\!\!\!\int 
d^{4}\hat{x} 
\left( \hat{\partial}^{+}_{A} 
\Phi_{1}(\hat{x})\right)
\Phi_{2}(\hat{x})\, .
\end{equation}
It is easy to check that
\begin{eqnarray}
 \hat{\partial}^{+}_{A}
\Phi(\hat{x}) =&&
{1\over (2\pi)^{4}}
\int d^{4}p \, 
e^{- {3p_{0}/ \kappa}} \chi_{A}
\cr\cr
&& \times
\left( - e^{p_{0}/ \kappa}
\vec{p}, -p_{0}\right)
\widetilde{\Phi}(p):e^{ip\hat{x}}:
\end{eqnarray}
where the functions $\chi_{A}(p)$ are given by (2.10).

\section{$\kappa$--deformed $\lambda \phi^{4}$ theory}

In order to write $\kappa$--deformed KG free equation we observe
that (see (2.10), $\mu=0,1,2,3$)
\begin{equation}
\chi_{\mu}\chi^{\mu} =
M_{\kappa}^{2} (p) 
\left( 1 - {M_{\kappa}^{2}(p)\over 4\kappa^{2}} \right)\, ,
\end{equation}
We write the free $\kappa$--deformed KG action as follows:

\begin{eqnarray}
\displaystyle
S_{0} =& &
\int\!\!\!\!\!\int 
 d^{4} \hat{x}
\left[
\left( 
\hat{\partial}^{+}_{\mu} \Phi^{+}(\hat{x})\right)
\partial^{\mu} \Phi(\hat{x})
\right.
\cr\cr
&&\left.
- \  m^{2} \Phi^{+}(\hat{x}) \Phi(\hat{x}) \right]
\cr
\cr
\displaystyle
=&&
 \int\!\!\!\!\!\int
 d^{4} \hat{x} \Phi^{+}(\hat{x})
 \left( 
{ \widehat{\Box}}
  - m^{2}
\right)
\Phi(\hat{x})\, ,
\end{eqnarray}
where $
{ \widehat{\Box}}
 \equiv \hat{\partial}_{\mu}\hat{\partial}^{\mu}$,
and we assume the
 following interaction
\begin{equation}
S_{\rm int} = {\lambda\over 4}
\int\!\!\!\!\!\int d^{4} \hat{x} \left( \Phi^{+}(\hat{x})
\Phi(\hat{x})\right)^{2}\, .
\end{equation}
By adding to the fields $\Phi$, $\Phi^{+}$ classical
variations\footnote{We follow the technique applied e.g. to
quantized fields in [18], where the quantum fields have c-number
variations.} we obtain the following field equation:
\begin{equation}
\left(
{ \widehat{\Box}}
- m^{2}\right)
\Phi =
{\lambda \over 4}
\left[ \Phi \left(
\Phi^{+}\Phi\right) +
\left( \Phi^{+}\Phi\right)\Phi\right]\, .
\end{equation}
Further,  we shall assume for simplicity that 
$\Phi^{+}(\hat{x})=\Phi(\hat{x})$, i.e., we assume that 
$\widetilde{\Phi}^{+}(\vec{p},p_{0})=
\widetilde{\Phi}(\vec{p},p_{0})$ (see (2.19)). We obtain  for
real $\kappa$--deformed KG fields
\begin{eqnarray}
S_{0} &=& {1\over 2}\int\!\!\!\!\!\int d^{4} \hat{x}
\Phi(\hat{x})
\left(
{ \widehat{\Box}}
 -m^{2}\right)
 \hat{\Phi}(\hat{x}) 
 \\
 \cr
 \displaystyle
& = & {1\over 2} \int d^{4} p
 \widetilde{\Phi} (-p)\left( M^{2}(p)
 \left(1 - {M^{2}_{\kappa}(p)\over 4\kappa^{2}} \right)
  - m^{2}\right)
 \widetilde{\Phi}(p)
 \nonumber
 \end{eqnarray}
The classical $\kappa$--deformed free KG field (see (1.10) and
(3.2)) is described by the Lagrangean ($\widehat{M}^{2}_{\kappa} =
 - M^{2}_{\kappa}({1\over i} \vec{\nabla}, {1\over i} \partial_{0})$)
\begin{equation}
S_{0} = \int d^{4}x  \, \phi(x)
\left[ \widetilde{M}^{2}_{\kappa} \left(
1 + {\widetilde{M}^{2}_{\kappa} \over 4\kappa^{2}} \right)
- m^{2}\right] \phi(x)
\end{equation}
where
\begin{eqnarray}
 \widetilde{M}^{2}_{\kappa}
& =& \Delta 
  e^{- {i\partial_{0}\over \kappa}}
   - \left( 2\kappa \sin {\partial_{0}\over 2\kappa}\right)^{2}
   \cr\cr
& =& \Delta  e^{- {i\partial_{0}\over \kappa}} 
- 2\kappa^{2}\left( 1 - \cos {\partial_{t}\over \kappa}\right)
 \end{eqnarray}
 and the free field equation takes the form
 \begin{equation}
\left(  \widetilde{M}^{2}_{\kappa} - m^{2}_{\kappa +}
\right)
\left(
  \widetilde{M}^{2}_{\kappa} - m^{2}_{\kappa -}
\right)\phi = 0
\end{equation}
where $  m^{2}_{\kappa \pm}=- 2\kappa^{2}\left( 1\mp
\sqrt{1 +{m^{2}\over \kappa^{2}}} \right)$. i.e.
\begin{mathletters}
\begin{eqnarray}
&& m^{2}_{\kappa +} = m^{2} - { m^{4}\over 4\kappa^{2} } + 
O\left({1\over \kappa^{4}}
\right)
\\ \cr
&& m^{2}_{\kappa -} =- 4\kappa^{2}- m^{2} + {m^{4}\over 4\kappa^{2}} 
+ O\left({1\over {4}}
\right)
\end{eqnarray}
\end{mathletters}
We see that the first mass $m^{2}_{\kappa +}$ describes the physical
spectrum, while the second one $ m^{2}_{\kappa -}$
  represents the tachyonic regularization mass,
in the spirit of Pauli--Villars regularization. Indeed, writing
the causal propagator, corresponding to (3.8) we obtain

\begin{eqnarray}
&&
\Delta^{F}_{\kappa} (\vec{p}, p_{0}) =
\\ \cr
&& =
\left(M^{2}_{\kappa} \left(\vec{p},p_{0}\right)
\left(
 { M^{2}_{\kappa} \left(\vec{p},p_{0}\right)\over 4\kappa^{2}}
- 1 \right) - m^{2} + i\epsilon\right]^{-1}
\cr\cr
&&
= {4\kappa^{2}\over \left( M_{\kappa}^{2} (p)
 - m^{2}_{+}  +i\epsilon
\right) \left( M_{\kappa}^{2} (p)
 - m^{2}_{-}  +i\epsilon \right) }
\cr\cr
& &
  =  {1\over \sqrt{1+{m^{2}\over \kappa^{2}} } }
 \left( {1 \over M_{\kappa}^{2} (p) - m^{2}_{+}  +i\epsilon }
 -  {1 \over M_{\kappa}^{2} (p) - m^{2}_{-}+i\epsilon } \right)
 \nonumber
 \end{eqnarray}

Defining
\begin{eqnarray}
&&\Delta^{F(\pm)}_{\kappa}(\vec{x},x_{0})=
\\ \cr
&&=\,  {1\over (2\pi)^{4}}
 \int d^{3}\vec{p} e^{i\vec{p}\, \vec{\kappa}}
  \int\limits^{+\infty}_{-\infty} dp_{0}
 {e^{-ip_{0}x^{0}}\over 
 M_{\kappa}^{2} (p) - m^{2}_{\pm} +i\epsilon}
 \nonumber
 \end{eqnarray}
 we get

 \begin{eqnarray}
&&\Delta^{F}_{\kappa}(\vec{x},x_{0})=
\\ \cr
&&= {1\over (2\pi)^{4}}
 \int d^{3}{p} e^{i\vec{p}\, \vec{\kappa}}
 \int\limits^{+\infty}_{-\infty} dp_{0}
 e^{-ip_{0}x^{0}}
 \Delta^{F}_{\kappa} (\vec{p},p_{0}) 
 \cr\cr
 && = 
  {1\over \sqrt{1+{m^{2}\over \kappa^{2}} } } 
  \left(
  \Delta^{F(+)}_{\kappa}(\vec{x},x_{0})
  - \Delta^{F(-)}_{\kappa}(\vec{x},x_{0})
  \right)
  \nonumber
   \end{eqnarray}
 In order to calculate short distance behaviour of (3.11--12)
let us observe that 
 $( M_{\kappa}^{2} (p) - m^{2}_{\pm})^{-1}$ in (3.11) contains
the following double infinite sequence of poles:
\begin{eqnarray}
\Delta^{F(\pm)}: \, & p_{0;n}^{(\pm)} = & \kappa \ln 
\left( { \pm \sqrt{1+{m^{2}\over \kappa^{2}} } +
 \sqrt{ {\vec{p}\, ^{2}+ m^{2}\over \kappa^{2} } } \over
   {p^{2}\over \kappa^{2}}  -1 } \right)
   \cr\cr
&&   +\,  2\pi i n\kappa =p_{0}^{(\pm )}+2\pi i n\kappa
   \cr\cr
   &
   p_{0;n}^{\prime(\pm )}  =  &\kappa \ln 
\left( {  \sqrt{1+{m^{2}\over \kappa^{2}} } \mp
 \sqrt{ {p}^{2}+ m^{2}\over \kappa^{2} }  \over
   {p^{2}\over \kappa^{2}}  -1 } \right)
\cr\cr
&&
   +\,  2\pi i\left( n + {1\over 2} \right) \kappa 
   \cr\cr
      & 
      \phantom{   p_{0;n}^{\prime(+)} }
       = & p_{0}^{\prime(\pm )} +2\pi i( n + {1\over 2} )\kappa
   \end{eqnarray}
   Calculating infinite summ of contributions from residua, e.g.

\begin{equation}
\sum\limits^{\infty}_{n=0} e^{-ix^{0}(p_{0}^{(+ )} +2i\pi n \kappa)}
= {e^{-i\kappa x_{0} p_{0}^{(+)}}\over 
\left( 1 - e^{2\pi \kappa x^{0}}\right)}
\end{equation}
and using the relation
\begin{equation}
\int\limits^{\infty}_{0} dp\, e^{ip|\vec{x}|}
\left({p\over \kappa}\right)^{i\kappa x_{0}} = 
\kappa \left( -i\kappa |\vec{x}| \right)^{-1-i\kappa x^{0}}
\Gamma\left(1+i\kappa x^{0}\right)
\end{equation}
one gets, after some tedious calculations, 
 for small $|\vec{x}|$, $ x_{0}$ the formula
\begin{equation}
\Delta_{\kappa}(\vec{x},x_{0}) \sim
{-i\kappa \cosh \left( {\pi \kappa x^{0} \over 2}\right) \Gamma  
\left( 1 + i \kappa x^{0}\right) \over
\left(2\pi\right)^{2} \sqrt{1 + {m^{2}\over \kappa^{2}} }\,
|\vec{x}| \left( \kappa | \vec{x}|\right)^{1+ i \kappa x^{0} } }
\end{equation}

In order to calculate the corrections due to interactions we
should rewrite the interaction term (3.3) in momentum space. For
real $\kappa$--deformed scalar field one gets

 \begin{eqnarray}
 \displaystyle
 S_{\rm int}&=& {\lambda\over 4}
 \int\!\!\!\!\!\int d^{4}\hat{x} \Phi^{4} (\hat{x})
 = {\lambda\over 4}
 \int d^{4}p^{(1)} \ldots d^{4}p^{(4)}
 \cr
 \cr
 \displaystyle
&&  \times \  \widetilde{\Phi}(p^{(1)})   \ldots
 \widetilde{\Phi} (p^{(4)})\delta^{4}
 \cr\cr
&& \times \ \delta
 \left(
 \Delta_{\mu}^{(4)}(p^{(1)},p^{(2)},p^{(3)},p^{(4)}
 \right)\, ,
 \end{eqnarray}
 where\footnote{It should be pointed out that the coassociativity
of the coproduct $\Delta^{(4)}$  provides the associativity of
the product $\Phi^{4}(\hat{x})$, i.e.,
$\Phi(\hat{x})\Phi^{2}(\hat{x}) =
\Phi^{2}(\hat{x})\Phi(\hat{x})$ etc.} (cf. (1.12c))

 \begin{eqnarray}
 \displaystyle
 \Delta^{(4)}_{0} &=& p^{(1)}_{0} +p^{(2)}_{0} + p^{(3)}_{0}  +
p^{(4)}_{0}\, ,
 \cr\cr
 \displaystyle
  \Delta^{(4)}_{i} &=&
  \sum\limits^{4}_{k=1}  p^{(k)}_{i}
  e^{- {1/ \kappa} \sum\limits^{4}_{l=k+1} p^{(l)}_{0}}\, .
  \end{eqnarray}
  One gets further
  
  \begin{eqnarray}
  S_{\rm int}& =& {1\over 4}
  \int 
   d^{4}   p^{(1)} 
    d^{4}   p^{(2)} 
     d^{4}   p^{(3)}
     \widetilde{\Phi}
     \left( \vec{p}\, ^{(1)},p^{(1)}_{0}
     \right)
     \cr\cr
     && \times
     \widetilde{\Phi}
     \left( \vec{p}\, ^{(2)},p^{(2)}_{0}
     \right)
     \widetilde{\Phi}
     \left( \vec{p}\, ^{(3)},p^{(3)}_{0}
     \right)
     \cr
    \cr
  &&  \times
    \widetilde{\Phi}
    \left(
    -\vec{p}\, ^{(1)}e^{p^{(1)}_{0}\over \kappa}
    - 
    \vec{p}\, ^{(2)}e^{ p^{(1)}_{0}+p^{(2)}_{0}\over \kappa}
\right.
\cr\cr
&&
\left.
-\ \vec{p}\, ^{(3)}e^{ p^{(1)}_{0}+p^{(2)}_{0}+ p^{(3)}_{0}\over \kappa},
- p^{(1)}_{0} - p^{(2)}_{0}- p^{(3)}_{0}\right)
\end{eqnarray}
In order to describe the Fourier transform as a nonlocal
expression in standard Minkowski space we change the variables
\begin{equation}
\vec{q}\, ^{(1)} =
e^{- {(p^{(2)}_{0} +p^{(3)}_{0})\over \kappa}} 
\vec{p}\, ^{(1)}
\quad
\vec{q}\, ^{(2)} =
e^{- {p^{(3)}_{0}/  \kappa}} 
\vec{p}\, ^{(2)}
\quad
\vec{q}\, ^{(3)} =
\vec{p}^{(3)}
\end{equation}
and put $q^{(i)}_{0}=p^{(i)}_{0}$. One obtains

\end{multicols}
{\hrule width0.5\textwidth}
\begin{eqnarray}
S_{\rm int} = &&
\displaystyle
 {\lambda    \over 4}
\int d^{4}q^{(1)} 
d^{4}q^{(2)}
d^{4}q^{(3)} 
e^{{3/ \kappa} 
\left( 2 q^{(1)}_{0}+q^{(2)}_{0}\right)}
\widetilde{\Phi}
\left(
\vec{q}\, ^{(1)}
e^{\left( {q}^{(2)}_{0}+{q}^{(3)}_{0} \right)\big/ \kappa}
, {q}^{(1)}_{0}
\right)
  \widetilde{\Phi}
\left(
\vec{q}\, ^{(2)} e^{q_{0}/ \kappa}, 
{q}^{(2)}_{0}
\right)
\widetilde{\Phi}\left(
\vec{q}\, ^{(3)}, {q}^{(3)}_{0}
\right)
\cr\cr
&& \times
\widetilde{\Phi}
\left( -
\left(
\vec{q}\, ^{(1)} + \vec{q}\, ^{(2)} +\vec{q}\, ^{(3)}
\right)
\displaystyle
e^{ \left( {q}^{(1)}_{0}+ {q}^{(2)}_{0} +{q}^{(3)}_{0} \right)\big/ \kappa },
- \left(
{q}^{(1)}_{0}+{q}^{(2)}_{0}+{q}^{(3)}_{0} \right)\right)
\cr\cr
 = && {\lambda \over 4} 
\int d^{4} x \phi^{4}_{\kappa} (x)
\end{eqnarray}
where

\begin{eqnarray}
\displaystyle
\phi^{4}_{\kappa} (x) = &&
e^{ {1/ \kappa} 
\left[ \left(
\partial^{(2)}_{0}+\partial^{(3)}_{0} \right)
\vec{x}\, ^{(1)}\vec{\nabla}\, ^{(1)}
+
\partial^{(3)}_{0} 
\left( 
\vec{x}\, ^{(2)}\vec{\nabla}\, ^{(2)}\right)
+
\partial^{(4)}_{0} 
\left( 
\vec{x}\, ^{(4)}\vec{\nabla}\, ^{(4)}\right)
\right]}
\cr
\cr
\displaystyle
&& \times 
\Phi
\left( \vec{x}\, ^{(1)}, x^{(1)}_{0} - {6i\over \kappa} \right)
\Phi
\left( \vec{x}\, ^{(2)}, x^{(2)}_{0} - {3i\over \kappa} \right)
\Phi
\left( \vec{x}\, ^{(3)}, x^{(3)}_{0} \right)
\Phi
\left( \vec{x}\, ^{(4)}, x^{(4)}_{0} \right)
\Bigg|_{{x^{(i)}=x\atop
i=1,2,3,4}}
\end{eqnarray}
describes the nonlocal $\kappa$--deformed $\lambda \phi^{4}$ vertex.
If we restrict ourselves to the leading term linear in ${1\over
\kappa}$ we get
$S_{\rm int} = S^{(0)}_{\rm int}+ {1\over \kappa} S^{(1)}_{\rm
int} + \ldots $ where

\begin{eqnarray}
\displaystyle
S^{(1)}_{\rm int} =&&
\displaystyle
{\lambda\over 4}
 \int
  d^{4}p^{(1)}
   d^{4}p^{(2)} d^{4}p^{(3)} 
   \displaystyle
   \left( 
   p^{(1)}_{0}\vec{p}\, ^{(1)} 
   +
\left(
   p^{(1)}_{0} + p^{(2)}_{0}\right)
   \vec{p}\, ^{(2)}
   +
   \left(
      p^{(1)}_{0} +    p^{(2)}_{0} +    p^{(3)}_{0}
\right)
\vec{p}\, ^{(3)}\right)
\cr\cr
&&\times 
\displaystyle
\widetilde{\Phi}
\left( \vec{p}\, ^{(1)},  {p}^{(1)}_{0}\right)
\widetilde{\Phi}
\left( \vec{p}\, ^{(2)},  {p}^{(2)}_{0}\right)
\widetilde{\Phi}
\left( \vec{p}\, ^{(3)},  {p}^{(3)}_{0}\right)
{\partial\over \partial \vec{p}}
\widetilde{\Phi}
\left( \vec{p};-  {p}^{(1)}_{0} -  {p}^{(2)}_{0} -  {p}^{(3)}_{0}\right)
\Bigg|_{\vec{p}=-\vec{p}_{1}-\vec{p}_{2}-\vec{p}_{3}}\, ,
\end{eqnarray}
i.e.,  one can write

\begin{eqnarray}
\displaystyle
\phi^{4}_{\kappa}(x)
&
\displaystyle
= &\phi^{4}(x) +
{\vec{x}\over \kappa}
\left(
\partial^{(1)}_{0}\vec{\nabla}\, ^{(1)}
+ \left( \partial^{(1)}_{0}+\partial^{(2)}_{0}\right)
\vec{\nabla}\, ^{(2)} 
+ \, \left( \partial^{(1)}_{0}+ \partial^{(2)}_{0} +\partial^{(3)}_{0}
\right)
\vec{\nabla}\, ^{(3)}\right)
\cr\cr
&&\times 
\Phi\left( \vec{x}\, ^{(1)}, x^{(1)}_{0}\right)
\Phi \left( \vec{x}\, ^{(2)},x^{(2)}_{0} \right)
\displaystyle
\Phi \left( \vec{x}\, ^{(3)},x^{(3)}_{0}\right)
\Phi\left(\vec{x}\, ^{(4)},x^{(4)}_{0}\right)
\Bigg|_{{x^{(i)}=x\atop
i=1,2,3,4}}
 + O({1\over \kappa^{2}}) \, .
\end{eqnarray}
\begin{multicols}{2}
It should be mentioned, however, that the expansion
 in 
$1\over \kappa$ provides the terms proportional to the powers of
$\vec{x}$ and the nature of the nonlocality described by the
exponential of differential operator is 
  not well described
  by such a power
expansion.\footnote{In similar way the finite
shift operator $e^{a\partial_{x}} $ is very inaccurately
approximated by the powers $(a\partial_{x})^{n}$.}

The formulae (3.21) and (3.23) describing nonlocal interaction
vertex can be used in the case when the field  $\Phi(x)$ is
classical (the Fourier components $\widetilde{\Phi}(x)$  commuting)
as well as in the quantized case.
Using perturbative  expansion in coupling constant $\lambda$ one
can derive
from (3.12) and (3.21) the modified rules for $\kappa$--deformed
Feynman diagrams, with

i) the internal lines 
  described by the propagator (3.12)

ii) the vertices introducing $\kappa$--deformed four--momentum
conservation law (see (3.13))
\begin{equation}
\delta^{4}\left( p_{1} + \ldots  + p_{4}\right)  \to
\Delta^{(4)}\left(p_{1}, p_{2},p_{3},p_{4}\right)\, .
\end{equation}
Unfortunately due to (3.2a) the $\kappa$--deformed Feynman diagrams will have
an unusual property:
the fourmomentum through virtual lines is not conserved. For
example if we consider the self--energy diagram in $\kappa$--deformed
 $\lambda \phi^{4}$ theory 
  the standard undeformed formula 
 \begin{figure}[h]
 \begin{minipage}[t]{8cm}
\epsfig{file=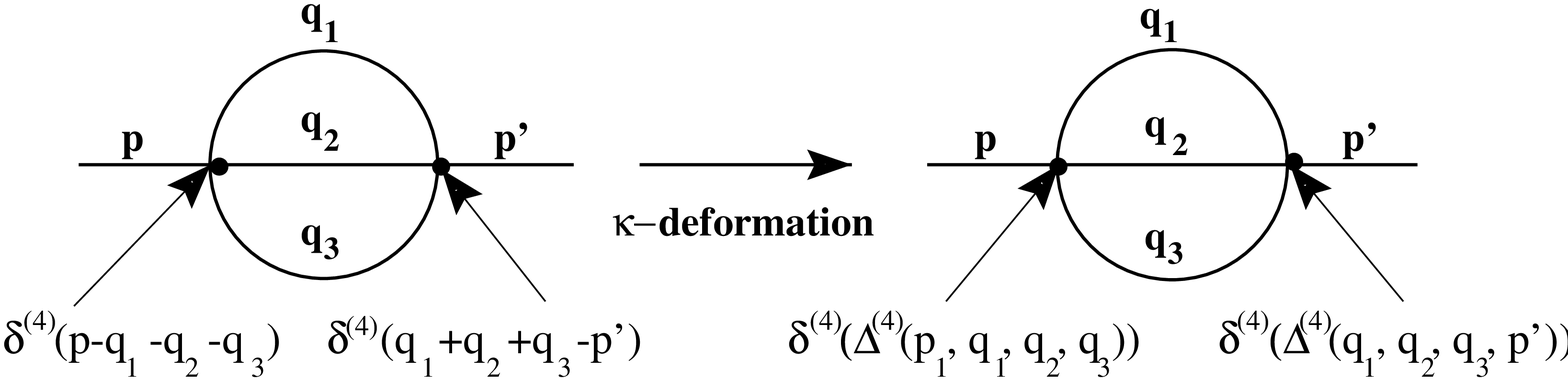,width=8cm}
\caption{ $\kappa$--deformation  of self--energy diagram  in
$\lambda \phi^{4}$ theory.}
\end{minipage}
\end{figure}
 it is easy to show that
 \begin{eqnarray}
&& \int d^{4}q_{1}\, d^{4}q_{2}\,d^{4}q_{3}\,  
F(q_{1},q_{2},q_{3})
\delta^{4} \left(p-q_{1}-q_{2}-q_{3}\right)
\cr\cr
&& \qquad \cdot
\delta^{4} \left( q_{1}+q_{2}+q_{3}-p^{\prime}\right)\sim
\delta^{4} \left(p-p^{\prime}\right)
\end{eqnarray}
is replaced by
 \begin{eqnarray}
&& \int d^{4}q_{1}\, d^{4}q_{2}\,d^{4}q_{3}\,  
 F_{\kappa}\left(
 q_{1},q_{2},q_{3}\right)
\cr\cr
&&\qquad \cdot
 \delta^{4}\left(\Delta^{(4)}\left(
 p,q_{1},q_{2},q_{3}\right)\right)
 \cr\cr
 && \qquad \cdot
  \delta^{4}\left(\Delta^{(4)}\left(
 q_{1},q_{2},q_{3},p^{\prime}\right)\right)
 \not\,\not{\!\!\sim}
 \delta^{4}\left(p-p^{\prime}\right)
 \end{eqnarray}
 We see from (3.27) that 
 the virtual process imply the dissipation of the fourmomenta.

At present we consider this nonconservation of the fourmomenta
at $\kappa$--deformed vertices as a serious difficulty.
\section{Conclusions}
We would like to point out that in this paper we present a
new scheme providing the rules how  to calculate the corrections to the
fourdimensional local 
 relativistic interacting  field theory in the presence of quantum
deformations\footnote{It should be mentioned that the deformation
of local interaction vertices in the presence of space--time coordinates
commuting to a nonvanishing c--number (see [1]) were considered by Filk [22].
Recently also the deformations of two--dimensional theories were
considered in [23].}. Our scheme can be described by the following diagram:
\begin{figure}[h]
\epsfig{file=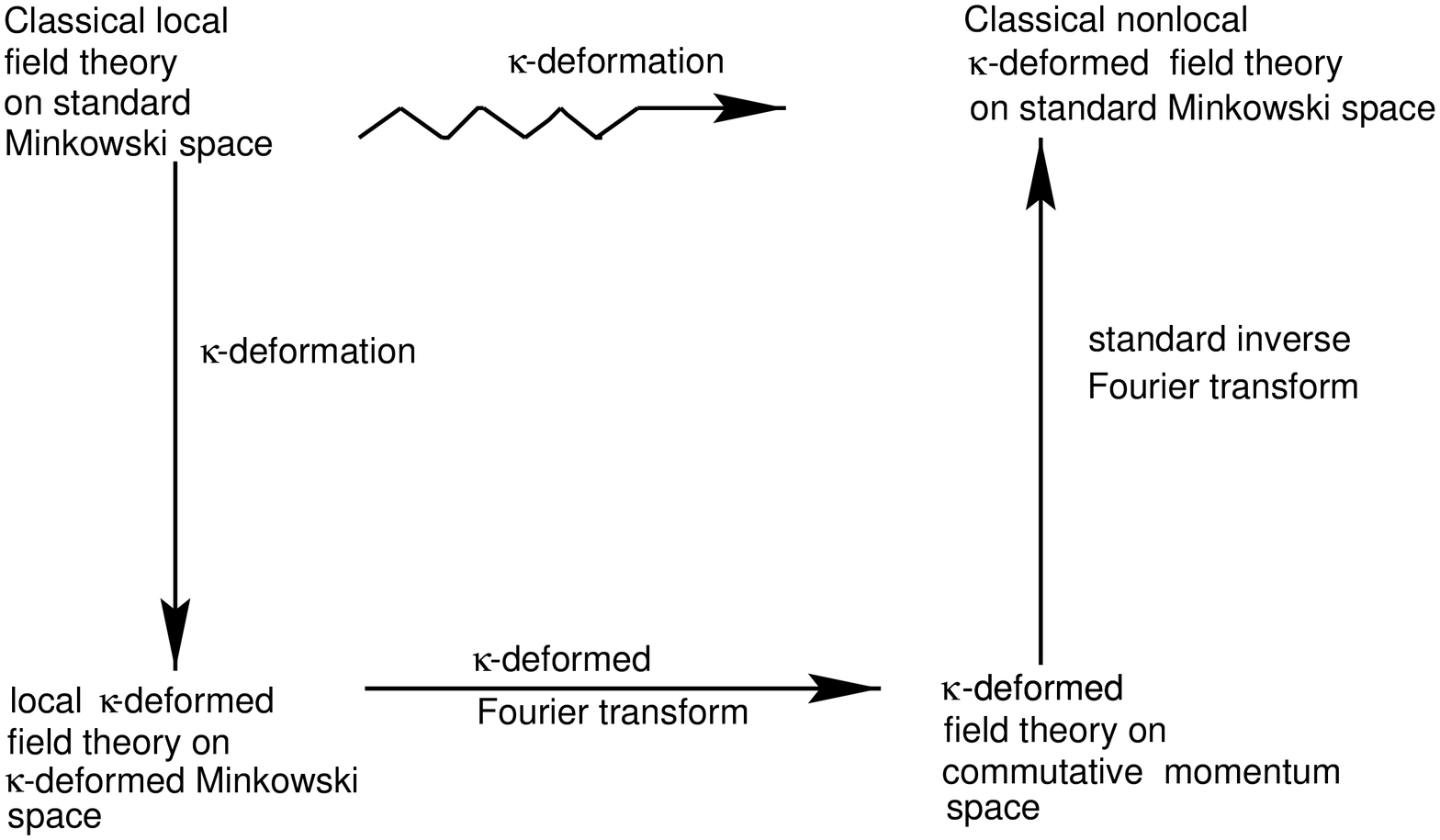, width=8cm}
\end{figure}

The scheme is valid for any deformation of D=4 Poincar\'{e}
symmetries with commutative fourmomenta. Such deformations can
be described by the Poincar\'{e} quantum group with the
following bicrossproduct structure:
\begin{equation}
{\cal P}^{(q)} = O(3,1) \ltimes T_{4}^{(q)}\, ,
\end{equation}
where

- $O(3,1)$ is the classical Lorentz algebra, with primitive
coproducts (classical Hopf-Lie algebra).

- $T^{(q)}_{4}$ describes the Hopf algebra of translations (see
(1.5a-b) in our case) deformed in a way 
preserving the primitive coproduct (1.5b).

It appears that the deformed space--time translations 
   $T^{(q)}_{4}$ should be
described by a set of relations
$$
\left[ \hat{x}_{\mu}, \hat{x}_{\nu}\right]=
{1\over \kappa} C_{\mu\nu}^{\rho}
\hat{x}_{\rho} + {1\over \kappa^{2}} T_{\mu\nu}\, ,
\eqno(4.2)
$$
with suitable conditions for the dimensionless coefficients
$C_{\mu\nu}^{\rho}$,
 $T_{\mu\nu}$, which can be obtained from the results presented
in [7]. The $\kappa$--deformation described in this paper is
distinguished by the property that it preserves the classical 
nonrelativistic $O(3)$ symmetries.

In conclusion we would like to point out the following two
problems which
we found while considering our corrections:

i) The modification of classical fourmomentum conservation law.
 
  The corrections to the local vertices
  are nonlocal in time and  from the 
 point of view of classical relativistic invariance are not
translation--invariant (see e.g. (3.21) and (3.24)). This
property occurs 
 always when in (4.2) $ C_{\mu\nu}^{\rho}\neq 0$\footnote{The
case of (4.2) with $ C_{\mu\nu}^{\rho}=0$ was considered in [1].
In such a case (see [22]) the classical fourmomentum
conservation at the vertices of deformed Feynman diagrams is valid.}

ii) The nonsymmetric coproduct of three--momenta requires some
novel approach to the problem of statistics and the notion of
bosons and fermions in the presence of $\kappa$--deformation.

The first difficulty means that the $\kappa$--deformed field
theory looks quite different from the undeformed one, 
and it is not yet clear how such modified
field theory
and modified energy--momentum conservation laws 
 could be useful in fundamental interactions theory.
This paper should be treated rather as an indication of 
 new features and 
 problems
which are met when we modify local field theory in a way covariant
under quantum--deformed Poincar\'{e} symmetries. In particular
because the Abelian addition law for the fourmomenta described
by their coproducts are modified (see (1.6b) and (1.12c) the
nonlocalities are necessarily required by the $\kappa$--deformed
form of translational invariance.

The second difficulty we consider to be rather of technical nature,
which should be solved by the introduction of some highly
nontrivial unitary operator, representing the exchange of
space--time position of two $\kappa$--deformed bosonic or
fermionic particles (for the solution of an analogous but much
simpler problem see [24]).

It should be mentioned that preliminary results of this paper have been 
presented at XXII International Colloquium on Group--Theoretic Methods
in July 1998 (Hobart, Tasmania).
\subsection*{Acknowledgments}
   The authors would like to thank Chrissomalis Chryssomalakos and Shahn Majid
for useful discussions. One of the authors (J.L.) would like to thank Jose
de Azcarraga and Departamento de Fisica for hospitality and the University of
 Valencia for financial support.

\section*{Appendix}
\renewcommand\theequation{A.\arabic{equation}}
\setcounter{equation}{0}

After tedious calculations one can show that
\begin{eqnarray}
&& :\, e^{-ip_{\mu}\left({\hat{x}}^{\nu}\otimes \Lambda_{\nu}^{\ \mu}
- I\otimes a^{\nu}\Lambda_{\nu}^{\ \mu}\right) }: 
\cr\cr
&&  = 
 e^{ip_{0}\otimes a^{\mu}\Lambda_{\mu}^{\ 0}}
  e^{ip_{k}\otimes a^{\mu}\Lambda_{\mu}^{\ k}}\, :
e^{-i \left( {\hat{x}}^{\mu} \otimes \bbbone \right)
\phi_{\mu}(p,\Lambda) }\, ,
\label{A.1}
\end{eqnarray}
where

$$
\phi_{0}(p,\Lambda) 
\displaystyle
=   \displaystyle \kappa \ln \Bigg[
\left\{ \cosh\,  {p_{0}\over \kappa} \otimes \bbbone 
+ \left( \sinh \, {p_{0}\over \kappa} \otimes \bbbone \right)
\circ H_{0}^{\ 0} \right\}
$$
$$
\phantom{\phi_{0}(p,\Lambda) =  }
  \times \left({\vec{p}\, ^{\, 2}\over 2\kappa^{2}} \otimes
\left( \Lambda_{0}^{\ 0} - 1\right)
+ {p_{\kappa}\over \kappa} \otimes \Lambda_{0}^{\ \kappa}
+ \bbbone \otimes \bbbone \right)\Bigg]\, ,
\label{A.2a}
\eqno(A.2a)
$$

\end{multicols}

$$
\displaystyle
 \phi_{k}(p,\Lambda) =
 {\kappa
\left(\sinh\,  {p_{0}\over \kappa} \otimes \bbbone \right)\circ
H_{k}^{\ 0}
\over
\left\{\cosh\,  {p_{0}\over \kappa} \otimes \bbbone
+ \left(\sinh \, {p_{0}\over \kappa} \otimes \bbbone \right)\circ
H_{0}^{\ 0} \right\}
\left( {p^{2}\over 2\kappa^{2}} \otimes \left(\Lambda_{0}^{\ 0}
- 1\right)
+ 2{p_{l}\over \kappa} \otimes \Lambda_{0}^{\ l} 
+ 2 \otimes \bbbone\right\} }\hfill
\eqno(A.2b)
\label{A.2b}
$$

\begin{multicols}{2}
\renewcommand\theequation{A.\arabic{equation}}
\setcounter{equation}{2}

and

$$
H_{0}^{\ 0} =
{2 
\otimes \Lambda_{0}^{\ 0} +{\vec{p}\, ^{\ 2}\over \kappa^{2}}
\otimes \left(\Lambda_{0}^{\ 0} - \bbbone\right)
+ 2{p_{k}\over \kappa} \otimes \Lambda_{0}^{\ k}\over
{\vec{p}\, ^{\ 2}\over \kappa^{2}}
\otimes \left(\Lambda_{0}^{\ 0} - \bbbone\right)
+ 2{p_{k}\over \kappa} \otimes \Lambda_{0}^{\ k}
+2\otimes \bbbone }
\eqno(A.3a)
$$

$$
H_{k}^{\ 0} =
{- 2{p_{l}\over \kappa}\otimes \Lambda_{k}^{\ l} +
 \left(\Lambda_{0}^{\ 0} - \bbbone\right)
+ 2{p_{l}\over \kappa} \otimes \Lambda_{0}^{\ l}
\Lambda_{k}^{\ 0}
+ 2 \otimes 
\Lambda_{k}^{\ 0}
\over
{\vec{p}\, ^{\ 2}\over \kappa^{2}}
\otimes \left(\Lambda_{0}^{\ 0} - \bbbone\right)
+ 2{p_{k}\over \kappa}
 \otimes \Lambda_{0}^{\ k}
+2\otimes \bbbone }\, .
\eqno(A.3b)
$$
\setcounter{equation}{3}
We obtain

\begin{eqnarray}
\Phi\left({\hat{x}}^{\nu}\otimes
 \Lambda_{\nu}^{\ \mu} \right. && 
 \left.
 -
\bbbone \otimes a^{\nu}
 \Lambda_{\nu}^{\ \mu} \right)
 \cr\cr  
= &&
 \int d^{4}p \ \widetilde{\Phi}(p)
 e^{ip_{0}\otimes a^{k}
  \Lambda_{k}^{\ 0}} 
     e^{ip_{k}\otimes a^{k}
  \Lambda_{k}^{\ k}} \, 
  \cr\cr
  &&
  :
    e^{-i({\hat{x}}^{\mu}\otimes \bbbone)
   \phi_{\mu}(p,\Lambda)}
\end{eqnarray}
  Using the  explicit formulae (A.2--3) and commutativity
  of $x^{\mu}$ with $\phi_{\mu}$ one gets
  
  \begin{eqnarray}
  \int\!\!\!\!\!\int d^{4} {\hat{x}} \, :
  e^{-i \hat{x}\phi(p,\Lambda)\otimes \bbbone} \, : 
= &&
  \left(2\pi \right)^{4} \delta^{4} \left(
  \phi(p,\Lambda) \right)\otimes \bbbone 
  \cr\cr
  = &&
  (2\pi)^{4}
  J\left({\partial\phi_{\mu}\over\partial p_{\nu}}\right)
  \delta^{4}(p) \otimes \bbbone
  \end{eqnarray}
  Because ${\partial \phi_{\mu}\over \partial
p_{\nu}} |_{p_{\rho}=0} =1 \otimes \Lambda_{\mu}^{\ \nu}$, we get
$J|_{p_{\rho}}=1$. Inserting (A.1) and (A.4) in (2.17) one
shows that the integral (2.16) has the same value for all frames
described by $\kappa$--deformed Poincar\'{e} group.

\end{multicols}

\begin{thebibliography}{99}
\bibitem{1} S.      Dopplicher, K. Fredenhagen and J. Roberts, Phys.
Lett. {\bf B331}, 39 (1994).
\bibitem{2} L.J. Garay, Int. Journ. Mod. Phys. {\bf A10}, 145 (1995).
\bibitem{3} A. Kempf and G. Magnano, Phys. Rev. {\bf D55}, 7909 (1997).
\bibitem{4} J. Lukierski, A. Nowicki, H. Ruegg and V.N. Tolstoy,
 Phys. Lett. {\bf B264}, 331 (1991).
 \bibitem{5} S. Zakrzewski, Journ. Phys. {\bf A27}, 2075 (1994).
 \bibitem{6} V. Dobrev, Journ. Phys. {\bf A26}, 1317 (1993).
 \bibitem{7}P. Podle\'{s} and S.L. Woronowicz, Comm. Math. Phys.
{\bf 178}, 61 (1996).
\bibitem{8}  S. Majid and H. Ruegg, Phys. Lett. {\bf B334}, 348 (1994).
\bibitem{9} S.L. Woronowicz, Comm. Math. Phys. {\bf 111}, 613 (1987).
\bibitem{10} C. Chryssomalakos,
 P. Schupp and P. Watts, preprint LBL-33274 (1992).
\bibitem{11} C. Fronsdal and A. Galindo, Lett. Math. Phys. {\bf
27}, 59 (1993).
\bibitem{12} F. Bonechi, E. Celeghini, R. Giachetti, C.M.
Pere\~{n}a,  E. Sorace and M. Tarlini, Firenze Univ. preprint
DFF 192/9/93.
\bibitem{13}S. Majid, Int. Journ. Mod. Phys. {\bf A5}, 4689 (1990).
\bibitem{14} S. Majid and R. Oeckl, math.QA/9811054.
\bibitem{15} S. Giller, C. Gonera, P. Kosi\'{n}ski, P. Ma\'{s}lanka, Journ.
Math. Phys. {\bf 37}, 5820 (1996).
\bibitem{16} C. Gonera, P. Kosi\'{n}ski, P. Ma\'{s}lanka, Acta
Phys. Polon. {\bf B27}, 2171 (1996).
\bibitem{17} C. Chryssomalakos and B. Zumino, Festschrift in
honour of A. Salam, Eds. A. Ali, J. Ellis, S. Ranjibar--Daemi (1993).
\bibitem{18} N.N. Bogolubov, V. Shirkov, ``Introduction to the
Theory of Quantum Fileds", Nauka, Moscov, 1973 (in Russian).
\bibitem{19}
 O.Ogievetsky, W.B. Shmidke, J. Wess and B. Zumino, Comm. Math. Phys. 
  {\bf 150},
495 (1992).
\bibitem{20}
J.A. de Azcarraga, P.P. Kulish and F. Rodenas, Fortschr. d.
Phys. {\bf 44}, 1 (1996).
\bibitem{21}
S.Majid, J. Math. Phys. {\bf 34}, 1176 (1993).
\bibitem{22}
T. Filk, Phys. Lett. {\bf B376}, 53 (1996).
\bibitem{23}
M. Chaichian, A. Demitchev and P. Presnajder, hep--th 9812180.
\bibitem{24}
P. Kosinski, P. Ma\'{s}lanka, ``Deformed Galilei symmetry"   
  math-9811142 (q-alg).
\end{thebibliography}
\end{document}